\documentclass[10pt]{revtex4}
                                                                                
                                                                                \usepackage{hyperref}
                                                                                \usepackage{amsmath,amsfonts,amssymb}
                                                                                \hypersetup{dvips,dvipdfm,colorlinks=true,urlcolor=magenta,filecolor=blue,linktoc=page,citecolor=red,linkcolor=blue,bookmarks=true}
                                                                                \usepackage{graphicx,epstopdf}
                                                                                
                                                                                \usepackage{array}
                                                                                \usepackage{makecell}
                                                                                \usepackage{float}
                                                                                \usepackage[thinlines]{easytable}
                                                                                \usepackage{multirow,bm}
                                                                                \usepackage{epstopdf}
                                                                                
                                                                                \newcounter{defcounter}
                                                                                \newcolumntype{L}[1]{>{\raggedright\let\newline\\\arraybackslash\hspace{0pt}}m{#1}}
                                                                                \newcolumntype{C}[1]{>{\centering\let\newline\\\arraybackslash\hspace{0pt}}m{#1}}
                                                                                \newcolumntype{R}[1]{>{\raggedleft\let\newline\\\arraybackslash\hspace{0pt}}m{#1}}

                                                                                \usepackage{color}

\begin{document}

\title{Cosmic scenarios in $f(R)$ gravity: a complete evolution
}

\author{Dipanjana Das$^1$\footnote {ddipanjana91@gmail.com}}
\author{Sourav Dutta$^2$\footnote {sduttaju@gmail.com}}
\author{Subenoy Chakraborty$^1$\footnote {schakraborty.math@gmail.com}}
\affiliation{$^1$Department of Mathematics, Jadavpur University, Kolkata-700032, West Bengal, India\\
	$^2$ Department of Pure Mathematics, University of Calcutta, 35, Ballygunge Circular Rd, Ballygunge, Kolkata-700019, West Bengal , India}


\begin{abstract}
The paper deals with $f(R)$ gravity theory in the background of inhomogeneous FLRW--type space time model. With proper choice of the inhomogeneous metric function it is possible to have an emergent scenario for the $f(R)$--cosmology. Explicit form of $f(R)$ is obtained for power law expansion. It has been shown that the present $f(R)$ gravity model is equivalent to some particle creation mechanism in Einstein gravity. Further a complete cosmic evolution from inflation to present late time acceleration is possible with proper continuous choices of the $f(R)$--functions. Finally, in the perspective of thermodynamical analysis a form of $f(R)$ has been evaluated using the unified first law.
	
\end{abstract}

\maketitle

\section{Introduction}
 The standard cosmology based on Einstein's general relativity (GR) and standard model of particle physics, can not accommodate various observational data (\cite{r1}-\cite{r5}) in recent years. Subsequently, by modification  of the particle physics the idea of exotic matter (known as dark energy), having large negative pressure has been introduced to give satisfactory explanations (\cite{r6}-\cite{r8}) to the above observations. On the other hand, extended gravity theories alternative to Einstein GR has been introduced to match with the observations (and also in the context of gravity quantization) (\cite{r9}-\cite{r11}). In this context, also from geometrical point of view $f(R)$ gravity theory (in which the Lagrangian density is an arbitrary function of the Ricci scalar $R$) is one of the simplest modifications. Also it is found that $f(R)$ gravity as well as 5D Dvali-Gabadadze-porrati (DGP) gravity (\cite{r12}-\cite{r13}) can produce the accelerated expansion of the Universe without introduction of any concept of dark energy (DE). Further, f(R)-gravity theory may resolve the hierarchy problem \cite{r14} and can explain the rotation curves of spiral galaxies without invoking any dark matter \cite{r15}, \cite{r16}. However, in the context of local tests in the solar system regime , several weak field constraints rule out many proposals for $f(R)$ gravity (\cite{r17}-\cite{r19}) though viable models do exist \cite{r20}, \cite{r21} (for details see review (\cite{r22}-\cite{r24}) on $f(R)$ gravity). Since Einstein's formulation of GR there are several attempts to modify or extend the theory but with quite disparate reasons. In order to unify electromagnetic and gravitational interaction geometrically Weyl \cite{r25} initiate such an attempt. Also motivated from quantum field theory in curved space--time there are modifications to the gravitational action for fundamental unification schemes as well as for a complete geometric explanation of the phenomenological \cite{r26} dark sector matters. f(R) gravity theory is possibly one of the simplest such extensions and inflation \cite{r27} is a natural consequence of this model. Recently, it has been shown \cite{r28} that $f(R)$ gravity theory can describe the cosmic evolution starting from initial inflation to the present late time acceleration without invoking any scalar field for accelerated expansion. The present paper is a similar type of work with $f(R)$--gravity theory to describe the continuous cosmic evolution from inflation to the present accelerated expansion. Also it has been shown that initial big--bang singularity may be avoided in this model and there is a phase of emergent scenario. The paper is organized as follows: Basic equations in $f(R)$ gravity theory for the present inhomogeneous model has been presented in section II. Emergent scenario and form of $f(R)$ for power law expansion shown in section III. Section IV shows the equivalence of the present model with the particle creation mechanism. A complete cosmic scenario has been described in section V for a continuous choice of the $f(R)$ function. Tests for $F(R)$ gravity models has been studied in section VI. A thermodynamical analysis using unified first law is presented in section VII. Finally, the paper ends with a short discussion in section VIII.
 \\

\section{Basic equations in $f(R)$ gravity theory and inhomogeneous FLRW--type model}

In $f(R)$ gravity, the action can be written as 
\begin{equation}
A=\frac{1}{2\kappa}\int \sqrt{-g}f(R) d^4x+\int d^4x \sqrt{-g} \mathcal{L}_m (g_{\mu \nu}, \psi),\label{e1}
\end{equation}
where $f(R)$ is an arbitrary function of the Ricci scalar $R$, $\mathcal{L}_m$ is matter Lagrangian density with metric tensor $ g_{\mu \nu}$ coupled to matter source denoted by $\psi$ and $\kappa=8\pi G$ is the usual gravitational coupling. Now variation of the above action with respect to the metric coefficients gives the fourth order field equations as 
\begin{equation}
F(R)R_{\mu \nu}-\frac{1}{2}f(R) g_{\mu \nu}-\bigtriangledown_\mu \bigtriangledown_\nu F(R)+g_{\mu \nu} \Box F(R)=\kappa T_{\mu \nu},\label{e2}
\end{equation}  
where 
\begin{equation}
T_{\mu \nu}=\frac{2}{\sqrt{-g}}\frac{\partial(\sqrt{-g} \mathcal{L}_{m})}{\partial g^{\mu \nu}},\label{e3}
\end{equation}
is the stress--energy tensor of the matter field, and $\Box=g^{\mu \nu}\bigtriangledown_\mu \bigtriangledown_\nu$ is the D'Alembertian operator and $F(R)=\frac{df(R)}{dR}$. Considering the trace of the field equations (\ref{e2}) we obtain
\begin{equation}
3\Box F(R)+R F(R)-2f(R)=\kappa T, \label{e4}
\end{equation}
which shows that $F(R)$ can be interpreted as a dynamical scalar degree of freedom of the theory with $T=T_{\mu}^{\mu}=g^{\mu \nu}T_{\mu \nu}$, the trace of the stress--energy tensor. Now using (\ref{e4}) in (\ref{e2}) after some algebraic manipulation one gets (with $\kappa=1$) \cite{r28.1}
\begin{equation}
G_{\mu \nu}=\frac{1}{F}T_{\mu \nu}+\frac{1}{F}\Bigg(\bigtriangledown_\mu \bigtriangledown_\nu F-g_{\mu \nu}N \Bigg)\label{e5}
\end{equation}
with 
\begin{equation}
N(t, r)=\frac{1}{4}\Bigg(RF+\Box F+T\Bigg).\label{e6}
\end{equation}

Although it is evident that the Universe is fundamentally not homogeneous on the scale of galaxies clusters and super--clusters there is clumping of matter, still homogeneous models of the Universe are widely used due to the elegance and simplicity of these models. However, at the very early era of evolution, the Universe was very likely to be in a state of much disorder (i.e., inhomogeneity). So it is interesting to have emergent scenario as well as inflationary era with inhomogeneous space-time model. Further, it is found that local inhomogeneities on account of their back reaction on the metric can produce apparent acceleration of the Universe \cite{a}, \cite{b}. Thus accelerated expansion may be due to inhomogeneous models of the Universe rather than introduction of dark energy. The simplest inhomogeneous toy modes corresponding to spherical symmetry are  (i) inhomogeneous FLRW model \cite{c, d} and (ii) Lemaitre-Tolman-Bondi (LTB) model \cite{f}--\cite{j} of the Universe.\\
 
The present work deals with spherically symmetric inhomogeneous (FLRW--type) space--time having line element
\begin{equation}
ds^2=-dt^2+a^2(t)\Bigg[\frac{dr^2}{1-\frac{b(r)}{r}}+r^2 d \Omega_2^2\Bigg],\label{e7}
\end{equation}
where $a(t)$ is the usual scale factor, $b(r)$ is an arbitrary function of $r$ and $d\Omega_2^2=d\theta^2+\sin^2 \theta d\phi^2$ is the metric on the unit 2--sphere.

Now the explicit form of the field equations (\ref{e5}) for the metric (\ref{e7}) can be written as
\begin{equation}
3H^2+\frac{b'}{a^2 r^2}=\frac{\rho}{F}+\frac{\rho_g(r, t)}{F},\label{e8}
\end{equation} 
\begin{equation}
-(2\dot{H}+3H^2)-\frac{b}{a^2 r^3}=\frac{p_r}{F}+\frac{p_{rg}}{F},\label{e9}
\end{equation}
and
\begin{equation}
-(2\dot{H}+3H^2)-\frac{b-rb'}{2a^2 r^3}=\frac{p_t}{F}+\frac{p_{tg}(r, t)}{F}.\label{e10}
\end{equation}
Here $\rho=\rho(r, t)$, $p_r=p_r(r, t)$ and $p_t=p_t(r, t)$ are the energy density, radial and transverse pressure of the cosmic anisotropic fluid under consideration. The conservation relations $(T^\nu_{;\nu}=0$) for this anisotropic matter are
\begin{equation}
\frac{\partial \rho}{\partial t}+H(3\rho+p_r+2p_t)=0,\label{e11}
\end{equation}
\begin{equation}
\frac{\partial p_r}{\partial r}=\frac{2}{r}(p_t-p_r).\label{e12}
\end{equation}
Similarly, $p_g, p_{rg}$ and $p_{tg}$ are the energy density and pressure components of a hypothetical curvature fluid having expressions:
\begin{eqnarray}
\rho_g&=&N+\ddot{F},\nonumber\\
p_{rg}&=&-N-H\dot{F}+\frac{r-b}{a^2r}F'-\frac{(b-rb')}{2a^2r^2}F',\nonumber\\
p_{tg}&=&-N-H\dot{F}+\frac{(r-b)}{a^2r^2}F'.\label{e13}
\end{eqnarray}
with an over dot and dash denoting differentiation with respect to cosmic time and radial coordinate respectively. Hence the field equations (\ref{e8})--(\ref{e10}) can be interpreted as the Einstein field equations with non-interacting two--fluids. Further the expression for scalar curvature for the present space--time model is
\begin{equation}
R=6(\dot{H}+2H^2)+\frac{2b'}{a^2 r^2}.\label{e14}
\end{equation}

\section{$f(R)$-gravity solutions and emergent scenario}

In order to obtain solutions for the $f(R)$ gravity cosmology we assume the inhomogeneous function $b(r)$ as
\begin{equation}
b(r)=b_0\big(\frac{r}{r_0}\big)^3+d_0,\label{e15}
\end{equation}
where $b_0, d_0$ and $r_0$ are constants.

This choice of $b(r)$ is such that the scalar curvature $R$ becomes a function of `$t$' alone:
\begin{equation}
R=6(\dot{H}+2H^2+\frac{b_0}{a^2 r_0^3}).\label{e16}
\end{equation}
Now using this choice of $b$ in the field equations (\ref{e8})-(\ref{e10}) we write the fluid components as
\begin{equation}
\rho=3H^2 \psi(t),~~~~~~~~~\rho_g=\frac{3b_0}{a^2 r_0^3},\label{e17}
\end{equation}
\begin{equation}
p_r=\psi(t)\Bigg\{-(2\dot{H}+3H^2)-\frac{d_0}{a^2 r^3}\Bigg\}-H\dot{\psi}(t),\label{e18}
\end{equation} 
\begin{equation}
p_t=\psi(t)\Bigg\{-(2\dot{H}+3H^2)+\frac{d_0}{2a^2 r^3}\Bigg\}-H\dot{\psi}(t),\label{e19}
\end{equation}
 \begin{equation}
 p_{rg}=p_{tg}=-\frac{b_0}{a^2 r_0^3}+H\dot{\psi}(t),\label{e20}
 \end{equation}
with $\psi(t)=F(R)$.

Thus the hypothetical curvature fluid is homogeneous and isotropic in nature while the cosmic fluid is chosen as inhomogeneous and anisotropic in nature. Also it is easy to see that the above choice of the matter distribution satisfy the conservation relations (\ref{e11}) and (\ref{e12}). Here the constant `$d_0$' is termed as inhomogeneous factor because for $d_0=0$ we have the usual FLRW model of the space--time with $b_0$ as the curvature scalar, and the model is inhomogeneous if $d_0\neq0$.\\

In the homogeneous case (i.e., $d_0=0$) if the cosmic fluid is assumed to be perfect fluid with $p=w\rho~(p_r=p_t=p)$ as the equation of the state then from equations (\ref{e17}) and (\ref{e18}) (or (\ref{e19})) we have
\begin{equation}
3(1+w)H=-\frac{2\dot{H}}{H}-\frac{\dot{\psi}( t)}{\psi(t)},\label{e21}
\end{equation}
which on integration gives 
\begin{equation}
\psi(t)=\frac{\psi_0}{\big[H^2\{exp\int3(1+w) \frac{da}{a}\}\big]},\label{e22}
\end{equation}
with $\psi_0$, a constant of integration.

So for constant equation of state  parameter
\begin{equation}
\psi(t)=\frac{\psi_0}{\big[H^2. a^{3(1+w)}\big]}.\label{e23}
\end{equation}
Assuming power law form of expansion of the Universe (i.e., $a=a_0t^n$) one gets
\begin{equation}
\psi(t)=\frac{\frac{\psi_0}{\big[n^2. a_0^{3(1+w)}\big]}}{t^{\{3nw+(3n-2)\}}}\nonumber
\end{equation}
and it can be expressed as a power law form of the scalar curvature for $n=1$ as
\begin{equation}
F(R)=\phi_0 R^{\frac{3w+1}{2}},\label{e24}
\end{equation}
and for $n=2$, we have
\begin{equation}
F(R)=\phi_1\Bigg\{\frac{R}{1-\sqrt{1+\xi R}}\Bigg\}^{3w+2}, \label{e24.1}
\end{equation}
with $\phi_0$ and $\phi_1$ are constants.

For the inhomogeneous case (i.e., $d_0 \neq 0$ if $w_r=\frac{p_r}{\rho}$ and $w_t=\frac{p_t}{\rho}$ are the equation of state parameters then from the equations (\ref{e17})--(\ref{e19}), they are linearly related as 
\begin{equation}
w_t-w_r=\frac{d_0}{2(a^2H^2)r^3}.\label{e25}
\end{equation}

Let us choose 
\begin{equation}
w_r=\frac{d_0}{6(a^2H^2)r^3}~ \mbox{and}~ w_t=\frac{-d_0}{3(a^2H^2)r^3},\label{e26}
\end{equation}

such that the conservation relations (\ref{e11}) and (\ref{e12}) are satisfied. Further, for this choice of equation of state using equations (\ref{e17})--(\ref{e19}) one gets the differential equation in $\psi$ as 
\begin{eqnarray}
\frac{\dot{\psi}}{\psi}+2\frac{\dot{H}}{H}+3H&=&0,\nonumber\\
\mbox{i.e.,} \psi&=&\frac{\psi_0}{(a^3 H^2)}.\label{e27}
\end{eqnarray} 
Also, as above one can obtain $\psi(t)$ as a power of scalar curvature if power law form of expansion of the Universe is assumed. On the otherhand, considering the hypothetical curvature fluid and comparing its energy density and thermodynamic pressure from the equations (\ref{e13}) and (\ref{e17}), (\ref{e20}) one obtains the differential equation in $\psi$ as 
\begin{equation}
\ddot{\psi}-2H\dot{\psi}-\frac{2b_0}{a^2r^3}\psi=0, \label{e28}
\end{equation} 
which has a simple solution
\begin{equation}
\psi=\frac{\psi_0}{2} \int \frac{d(a^2)}{H},\label{e29}
\end{equation}
for inhomogeneous model with $b_0=0$. So for the power--law expansion $F(R)$ takes the form $R^{-(n+\frac{1}{2})}$.

We shall now examine whether emergent scenario is possible for this inhomogeneous $f(R)$ gravity model. In emergent scenario the form of the scale factor should be of the form:
\begin{equation}
a=\big(a_0+a_1 e^{\mu t}\big)^\alpha,~~\mu, \alpha>0.\label{e30}
\end{equation} 
Then using this choice of $a(t)$ and the inhomogeneous function $b(r)$ from equation (\ref{e15}), the explicit form of differential fluid components are (for $d_0 \neq 0$)
\begin{equation}
\rho=\frac{3\psi_0}{a^3},~p_r=\frac{-d_0\psi_0}{a^5 H^2 r^3},~p_t=\frac{d_0\psi_0}{2a^5 H^2 r^3},\label{e31}
\end{equation}
(i.e., $w_r=-\frac{1}{3}$ and $w_t=\frac{1}{6}$) where the solution (\ref{e27}) for $\psi$ has been used. Note that 
$$\rho+p_r+2p_t=\frac{3\psi_0}{a^3}>0,$$
so the strong energy condition (SEC) is always satisfied. Thus for emergent scenario the inhomogeneous and anisotropic cosmic fluid under consideration satisfies the SEC i.e., non--exotic in nature. Further, for this emergent scenario $F(R)$ takes the form (for $\alpha$=1): 
\begin{equation}
F(R)=\psi_0 \frac{R^{3(w+1)}}{\Bigg[lR+\sqrt{m-zR-qR^2}\Bigg]^2\Bigg[uR+\sqrt{m-zR-qR^2}\Bigg]^{3w+1}},\label{e31.1}
\end{equation}
with $l, m, z, u$ and $q$ as a constant.\\

However, at emergent scenario the size of the Universe is so small (of the Planck order) that quantum field theory in curved space-time may be an adequate description. As a result the matter field must be quantized and we have the semi classical description of gravity as
$$G_{\mu \nu}=\big<T_{\mu \nu}\big>=\big<\psi|\hat{T}_{\mu \nu}\big|\psi\big>.$$
Here $\big|\psi\big>$ is a quantum state describing the early Universe and $\hat{T}_{\mu \nu}$ is the quantum operator associated with the classical energy-momentum tensor of the matter field. Thus due to this quantum formulation Einstein-Hilbert action is modified with non-linear invariants of the curvature tensor or non-minimal couplings between matter and curvature due to perturbative expansion \cite{a2}. Thus due to this quantum effect the very early phase of the Universe is affected and as a result emergent scenario may be considered due to this quantum effect (trace anomaly) and eliminate the big-bang singularity.

\section{Equivalence of f(R) gravity theory with particle creation mechanism }
This section is an attempt to show the equivalence between f(R)gravity theory and the particle creation mechanism in Einstein gravity. In non equilibrium thermochemical prescription the particle creation rate is related to the dissipative pressure for adiabatic process as \cite{r28.2}
\begin{equation}
\Gamma =-3H\frac{\Pi}{p+\rho},\label{d1}
\end{equation} 
where $\Gamma$ is the particle creation rate,$\Pi$ is the dissipative pressure and(p,$\rho$)are the usual energy density and thermodynamic pressure of the dissipative fluid.\\
We shall now consider the present f(R) gravity model both for homogeneous(i.e $d_0=0$)and inhomogeneous (i.e. $ d_0\neq0$) cases separately.\\

$\bullet$\underline{\bf Homogeneous Model:}$(d_0=0)$\\
In this case,we have:
$p_{rg}=p_{tg}$and $p=p_{r}=p_{t}$. So we assume that $p=\omega \rho$. Then from equation (\ref{e9}) (or (\ref{e10}))
$$-(2\dot{H}+3H^2)=\frac{p}{\psi(t)}+H\frac{\dot\psi}{\psi}.$$
So comparing with Einstein gravity the dissipative pressure is given by\\
$$\Pi=H\frac{\dot \psi}{\psi},$$
so from the equation(\ref{d1}),\\
$$\Gamma=\frac{-3H^2\frac{\dot \psi}{\psi}}{(1+\omega)\frac{\rho}{\psi}}.$$
Now using (\ref{e17}) for $\rho$ one gets
\begin{equation}
\psi(t)=\psi_{0}\exp\Bigg[-\int(1+\omega)\Gamma dt\Bigg].\label{d2}
\end{equation}

$\bullet$\underline{\bf Inhomogenous Model}:$( d_0\neq 0)$

In this case the real fluid is inhomogeneous and anisotropic in nature. So for the field equations (\ref{e9}) and (\ref{e10}) using (\ref{e20}) for $p_{rg}~ \mbox{and}~ p_{tg}$
$$-(2\dot{H}+3H^2)=\frac{p_r}{\psi}+\Bigg(H\frac{\dot\psi}{\psi}+\frac{d_0}{a^2 r^3}\Bigg)$$
$$-(2\dot{H}+3H^2)=\frac{p_t}{\psi}+\Bigg(H\frac{\dot\psi}{\psi}-\frac{d_0}{2a^2 r^3}\Bigg)$$
Hence the dissipative pressure is also anisotropic with
\begin{eqnarray}
\Pi_r= \Bigg(H\frac{\dot\psi}{\psi}+\frac{d_0}{a^2 r^3}\Bigg),
\Pi_t=\Bigg(H\frac{\dot\psi}{\psi}-\frac{d_0}{2a^2 r^3}\Bigg),\label{d3}
\end{eqnarray}
as the radial and tangential components. As a result, the particle creation rate is also anisotropic in nature and we have\\
\begin{eqnarray}
\Gamma_r&=&-(1+\omega_r)^{-1}~\frac{\dot \psi}{\psi}-\frac{d_0}{a^2 r^3H},\nonumber\\
\mbox{i.e.,}~ \psi&=&\psi_0~ \exp\Bigg[-\int\Bigg((1+\omega_r)\Gamma_r+\frac{(1+\omega_r)d_0}{a^2Hr^3}dt\Bigg)\Bigg],\label{d4}
\end{eqnarray}
and
\begin{eqnarray}
\Gamma_t&=&-(1+\omega_r)^{-1}~\frac{\dot \psi}{\psi}+\frac{d_0}{2a^2 r^3H},\nonumber\\
\mbox{i.e.,} \psi(t)&=&\psi_1 \exp\Bigg[\int\Bigg\{-(1+\omega_r)\Gamma_t+\frac{(1+\omega_t)d_0}{2a^2Hr^3}dt\Bigg\}\Bigg].\label{d5}
\end{eqnarray}
It is to be noted that in this case$(\Gamma_r,\Gamma_t)$ as well a $(\omega_r,\omega_t)$ are inhomogeneous (i.e. functions of $r$ and $t$)but $\psi$ in equations (\ref{d4}) and (\ref{d5}) is a function of ``$t$" alone and they should be equal. So for consistency one should have the following choices:
\begin{eqnarray}
1+w_r&=&w_{r_0}^{(t)}r^3,~1+w_t=w_{t_0}^{(t)}r^3,\nonumber\\
\Gamma_r&=&\frac{\Gamma_0(t)}{r^3},~\Gamma_t=\frac{\Gamma_1(t)}{r^3},\label{d6}
\end{eqnarray}
and hence from equations (\ref{d4}) and (\ref{d5})
\begin{equation}
\psi=\psi_0 \exp\Bigg[-\int \Bigg\{w_{r_0} \Gamma_0+\frac{w_{r_0}d_0}{a^2H}\Bigg\}dt\Bigg],\label{d7}
\end{equation}
and
\begin{equation}
\psi=\psi_1 \exp\Bigg[\int \Bigg\{-w_{t_0} \Gamma_0+\frac{w_{t_0}d_0}{2a^2H}\Bigg\}dt\Bigg].\label{d8}
\end{equation}
Now equality of these two expressions for $\psi$ demands
\begin{equation}
w_{r_0}=\frac{1}{2} w_{t_0},~\Gamma_0=-2\Gamma_1,~\psi_0=\psi_1.\label{d9}
\end{equation}
Hence, if there is particle creation along the radial direction then there should be particle annihilation along the transverse direction or vice--versa.

Before proceeding to the next section we shall present some explicit form of $\psi$ (in homogeneous case) for some choices for the particle creation rate from (\ref{d2}).

$\bullet$ $\Gamma=\alpha_0 t^n$,~$\alpha_0, n$ are constants from (\ref{d2})
\begin{eqnarray}
\psi(t)&=&\psi_0 \exp \Bigg[-\alpha_0 \frac{1+w}{1+n} t^{n+1}\Bigg],~\mbox{for}~n \neq -1\nonumber\\
&=&\psi_0 t^{(1+w)\alpha_0}~\mbox{for}~n=-1.\nonumber
\end{eqnarray}

$\bullet$ $\Gamma=\alpha_0 \exp(nt)$
\begin{equation}
\psi(t)=\psi_0 \exp\Bigg[\frac{-\alpha_0(1+w)}{n}e^{nt}\Bigg].
\end{equation}

\section{continuous cosmic evolution: complete scenario}

The aim of this section is to examine whether there exists a continuous choice of the function $F(R)$ so that the cosmic evolution from inflation to late time acceleration can be described in the present model. In an earlier work \cite{r28.2} it has been shown that the inflationary era, matter dominated era and late time accelerating phase can be described for three different choices of the particle creation parameter which is continuous across the transition points. Although, the choices of the creation rate parameter apparently seem to be phenomenological but there are thermodynamical arguments behind these choices of the particle creation parameter. \\

(i) $\Gamma=3\mu_1 \frac{H^2}{H_1}$ (inflationary era for $t \leq t_1$),

(ii)  $\Gamma=3\mu_2H,~t_1 \leq t \leq t_2$ (matter dominated era),

(iii) $\Gamma=3\mu_3 \frac{H_3}{H},~t\geq t_2$ (present accelerating phase).\\

In the previous section we have determined $\psi(t)$ in terms of $\Gamma$ for homogeneous case. Now the explicit form of $\psi(t)$ in these three different phases are given below:\\

{\bf Inflationary Era:}
\begin{eqnarray}
H&=&\frac{H_1}{\mu_1+(1-\mu_1)\Bigg(\frac{a}{a_1}\Bigg)^{\frac{3(1+w)}{2}}} ,\nonumber\\
\psi&=&\eta_1 \exp \Bigg[-\frac{2}{a_1} \ln \Bigg(1-\frac{\mu_1}{\mu_1+(1-\mu_1)(\frac{a}{a_1})^{\frac{3(1+w)}{2}}}\Bigg)\Bigg].\label{c1}
\end{eqnarray}

{\bf Matter dominated Era:}
\begin{eqnarray}
H&=&H_1\Bigg(\frac{a}{a_1}\Bigg)^{\frac{-3(1+w)(1-\mu_2)}{2}},\nonumber\\
\psi&=&\eta_2 \exp \Bigg[-3\mu_2 (1+w) \ln a\Bigg].\label{c2}
\end{eqnarray}

{\bf Late time accelerating phase:}
\begin{eqnarray}
H&=&\mu_3 H_2+\Bigg(\frac{a}{a_2}\Bigg)^{-3(1+w)},\nonumber\\
\psi&=&\eta_3 \exp \Bigg[-\frac{\mu_3 H_2}{a_2} \ln \Bigg(1-\frac{\mu_3H_2}{\mu_3 H_2+(\frac{a}{a_2})^{-3(1+w)}}\Bigg)\Bigg].\label{c3}
\end{eqnarray}
Here $(a_1, H_1)$ and $(a_2, H_2)$ are the values of the scale factor and Hubble parameter at the transition epochs $t=t_1$ and $t=t_2$ respectively and $\psi_1, \psi_2$ and $\psi_3$ are the constants of integration. Now for the continuity of $\psi$ at the junction point, the condition of this continuity  should be in the form of the following
\begin{eqnarray}
a_1 \ln a_1&=&\frac{2}{3} \frac{\ln (1-\mu_1)}{\mu_2 (1+w)},\nonumber\\
&\mbox{and}& \nonumber\\
a_2 \ln a_2&=&\frac{H_2^2}{1+w}\ln (1+\mu_3H_2),\label{c4}
\end{eqnarray}
provided $\eta_1=\eta_2=\eta_3$.

\section{Tests for $F(R)$ gravity models}
The admissibility of the present $f(R)$-gravity models will be discussed in this section. In fact, it will be examined whether these $f(R)$ models are consistent with existing experimental results as well as observed data.\\

In metric formulation, $f(R)$-gravity theory is known to be equivalent to scalar tensor gravity \cite{a3}--\cite{d3} with zero coupling parameter (i.e., $w=0$) \cite{c3} as in Brans-Dicke theory. As solar system tests do not allow small values of $w$ so these tests rule out $f(R)$ gravity models. However, the non-minimally coupled scalar degree of freedom in strongly massive region \cite{e3}--\cite{g3} can be largely suppressed due to acquisition of excess mass and hence such similarity between the two gravity theories may not always be justified. Further, any viable $f(R)$ gravity model should be consistent \cite{g3}--\cite{k3} with the tests by the following constraints:\\

I. $F(R)$ should be positive for all finite $R$ to keep the effective Newton's constant: $G_{eff}=\frac{G}{F(R)}$ to be positive. Moreover, at the microscopic level positivity of $G_{eff}$ does not allow the gravitons to be ghost particles \cite{k3}.\\

II. $\frac{dF}{dR}>0$ for large $R$ in the matter dominated Universe (i.e., for a stable high-curvature regime). So scalaron models do not become tachyonic at the quantum level.\\

III. $F(R)$ should be tightly constraint as $F(R)<1$, in big-bang nucleosynthesis and cosmic microwave background.\\

IV. Although the galaxy formation surveys restrict $F(R)$ to be $|F(R)-1|<10^{-6}$ \cite{r28.1} but this survey is not yet studied in $F(R)$ model using $N$-body simulation. So the above constraint is yet to be confirmed. \\

Thus for above constraints we can restrict the arbitrary parameters for our $F(R)$ solutions (\ref{e24}) and (\ref{e24.1}) as:

\begin{center}
	\begin{table}[!htb]
		\renewcommand{\arraystretch}{1.9}
		~~~~~~~\begin{tabular}{| >{\centering\arraybackslash}m{1cm}|>{\centering\arraybackslash}m{4cm}|>{\centering\arraybackslash}m{6cm}|}
			\hline
			Sl. no. & Solution of $F(R)$ & Parameter constraint \\
			\hline
			$1$ & $\phi_0 R^{\frac{3w+1}{2}}$ & $0<R<\Big(\frac{1}{\phi_0}\Big)^{\frac{2}{3w+1}}$\\
			& & $w>-\frac{1}{3},~\phi_0>0$ \\
			\hline
			$2$ & $\phi_1\Bigg\{\frac{R}{1-\sqrt{1+\xi R}}\Bigg\}^{3w+2}$ & $\phi_1 >0,~\xi>0, 3w~\mbox{is even}$\\
			& & $0< R < \mu(\xi \mu -2),~\mu=\Big(\frac{1}{\phi_0}\Big)^{\frac{1}{2+3w}}$\\ 
			\hline
		\end{tabular}
	\end{table}
\end{center}

\section{A thermodynamical analysis of the F(R) gravity theory in the present inhomogeneous model:}
The line element for the present inhomogeneous FLRW-type space-time (given in equation (\ref{e7})) can be expressed in terms of double null form as 
\begin{equation}
ds^2=-2 d \xi^{+} d \xi^{-}+R^2 d\Omega_2^2,\label{t1}
\end{equation}
where $d \xi^{\pm}=-\frac{1}{\sqrt{2}}\Bigg(dt \mp \frac{a dr}{\sqrt{1-\frac{b(r)}{r}}}\Bigg),$ are the null one-forms and $R=ar$ is the usual area radius. The space-time is assumed to be time-orientable with $\partial_{\pm}=\frac{\partial }{\partial \xi^{\pm}}=-\sqrt{2}\Bigg(\partial t \mp \frac{\sqrt{1-\frac{b(r)}{r}}}{a} \partial r\Bigg),$ as future pointing null vectors. Thus $\xi^{\pm}=$ constant represent two families of null geodesics and the expansion scalars associated with these congruences of null geodesics are
\begin{equation}
\theta_{\pm}=\frac{2}{R} \partial_{\pm} R.\label{t2}
\end{equation}
The positive or negative sign of the above scalars indicate the increase or decrease of the area of the sphere along the null directions.

According to Hayward \cite{r29}-\cite{r32}, a trapping horizon $(H_T)$ is a hypersurface in 4D space-time foliated by 2-spheres with the restrictions:
\begin{equation}
\theta_{+|H_T}=0, \theta_{-|H_T}\neq 0~\mbox{and}~ \mathcal{L}_{\theta_{+|H_T}}\neq 0.\label{t3}
\end{equation}
The above trapping horizon is identified as outer or inner according as $\mathcal{L}_{\theta_{+|H_T}}$ is positive or negative. Similarly, it is characterized as future or past depending on the sign $\theta_{-|H_T}$ as negative or positive. A dynamical black hole is defined \cite{r29}-\cite{r32} as the space-time region bounded by future outer trapping horizon.

In the present context, horizon will be similar to the cosmological horizon of the de--Sitter space-time. If $R_T$ is the area radius of the trapping horizon then $R_T$ is characterized by \cite{r29}-\cite{r34}
\begin{eqnarray}
\partial_{+} R\big|{_{R=R_T}}&=&0,\nonumber\\
\mbox{i.e.,}~R_T&=& \frac{1}{\sqrt{H^2+\frac{C(r)}{a^2}}}=R_A,\label{t4}
\end{eqnarray} 
with $C(r)=\frac{b(r)}{r}$.

As, $\partial_{-} R\big|_{{R=R_T}}=-2R_T H<0$, so the trapping horizon coincides with apparent horizon and is future in nature. Usually, in Universal thermodynamics any boundary of the Universe (analogous to BH horizon) is termed as horizon and in the present context trapping horizon is defined.

From the point of view of thermodynamical analysis the surface gravity and Hawking temperature on any horizon $(R=R_h)$ are defined as \cite{r35}
\begin{equation}
\kappa_h=-\frac{1}{2} \frac{\partial \chi}{\partial R}\Bigg|_{R=R_h}=\frac{R_h}{R_A^2},\label{t5}
\end{equation}
and
\begin{equation}
T_h=\frac{|\kappa_h|}{2\pi}=\frac{R_h}{2 \pi R_A^2},\label{t6}
\end{equation} 
with $\chi=h^{ab} \partial_a R \partial_b R$ ($h^{ab}$ is the 2-metric of the normal $(r, t)$-plane of the spherical symmetry).

Moreover, Hayward \cite{r29}-\cite{r32} has shown that for non-static spherically symmetric space-times Einstein field equation known as ``Unified first law" (UFL) which contains both physical and geometrical variables. The mathematical form of UFL is
\begin{equation}
dE=A\psi+WdV, \label{t7}
\end{equation}
where the one form $\psi$ represents energy flux (or energy supply), $W$ is the work density, $A, V$ are the area and volume bounded by the horizon and $E$ is the Misner--Sharp energy \cite{r29}-\cite{r34}. The explicit form of the above quantities are
\begin{eqnarray}
\psi&=&\psi_a dx^a, (a=0, 1),~~\psi_a=T_a^b \partial_b R+W \partial_a R,\nonumber\\
W&=&-\frac{1}{2} T_r. T=-\frac{1}{2}T^{ab} h_{ab},\nonumber\\
A&=&4\pi R^2,~V=\frac{4}{3} \pi R^3.\label{t8}
\end{eqnarray}
and
$$E=\frac{R}{2G}\Bigg(1-h^{ab} \partial_a R \partial_b R\Bigg).$$
Note that purely geometric quantity $E$ is related to the space-time structure and it measures the total energy inside the sphere of radius $r$. So for the present inhomogeneous model the above quantities take the form:
\begin{eqnarray}
\psi&=& \frac{1}{2}\big(\rho_t+p_{rt}\big)(-HRdt+adr),\nonumber\\
W&=&\frac{1}{2} \big(\rho_t-p_{rt}\big),\nonumber\\
E&=& 4\pi R^3 \Bigg(H^2+\frac{b}{rR^2}\Bigg)\label{t9}
\end{eqnarray}
with $\rho_t=\frac{1}{\psi(t)}\{\rho+\rho_g\},~p_{rt}=\frac{1}{\psi}\{\rho_r+\rho_{rg}\}.$

Now using (\ref{t9}) in the both sides of UFL in equation (\ref{t7}) one gets the Einstein field equations (\ref{e8}) and (\ref{e9}) on any horizon and this justifies the claim of Hayward.

Further, on any horizon, if Bekenstein's entropy--area relation is assumed and the temperature on the horizon is taken as the generalized Hawking temperature \cite{r36} i.e.,
\begin{equation}
T^{(g)}_h=\alpha T_h,\label{t10}
\end{equation}
then for the validity of the first law of thermodynamics (i.e., Clausius relation: $dE_h=T^{(g)}_h dS_h)$ one gets
\begin{equation}
\frac{\dot{E_h}}{\dot{R_h}}=\alpha|\kappa|R_h=\frac{E'_h}{R'_h},\label{t11}
\end{equation}
where $\alpha$ is a dimensionless parameter. The above relation (\ref{t11}) can be interpreted as the integrability condition for the following one form equation:
\begin{equation}
dE_h=\lambda dR_h,\label{t12}
\end{equation}
where $\lambda$ is an arbitrary (but smooth) function of $t$ and $r$.

On the other hand, Using Einstein field equations (\ref{e8}) and (\ref{e9}) into the above integrability condition (\ref{t11}) one gets (after simplification)
\begin{equation}
\rho_t+p_{rt}=0.\label{t13}
\end{equation}
 Thus for the validity of the Clausius relation there is no restriction on the parameter $\alpha$ rather it restricts the total effective fluid as a cosmological constant. Further using equations (\ref{e17})-(\ref{e20}) in relation (\ref{t13}) one determines $\psi(t)$ as
 \begin{equation}
 \psi(t)=e^{-\int3H(1+w_r)dt}\Bigg[\psi_0- \frac{2b_0}{r_0^3}\int \frac{dt}{a^2H} \exp \int 3H(1+w_r)dt\Bigg], \label{t14}
 \end{equation}
 with $\psi_0$, the constant of integration.
 
 Thus from the Clausius relation with the above integrability condition an explicit form of $\psi(t)$ is obtained.
 \section{Short Discussion} 
 The present work deals with cosmic evolution in $f(R)$-gravity theory in the background of a typical inhomogeneous space-time model known as inhomogeneous FLRW model. The justification of choosing inhomogeneous space-time model is given both from the point of view of early cosmic evolution as  well as from the present late time cosmic evolution. The inhomogeneous function $b(r)$ in the metric is chosen appropriately to make the scalar curvature homogeneous. As a result, there is a lot of simplification in the evolution equations. The functional form of $f(R)$ has been determined for power law expansion for some typical choices of the power parameter. The $f(R)$-gravity model has been shown to be equivalent to Einstein gravity in two ways. At first, non-interactions 2-fluid model in Einstein gravity has been shown to be equivalent to the present $f(R)$-gravity theory. Here the first fluid is the usual cosmic fluid which is inhomogeneous and anisotropic in nature while the hypothetical fluid (known as curvature fluid) is homogeneous and isotropic in character. Secondly, in the context of non-equilibrium thermodynamics $f(R)$-gravity theory can be considered as Einstein gravity with particle creation mechanism. Non-singular emergent scenario is possible for the present $f(R)$-gravity model with normal (i.e., non-exotic) matter. It has been shown that quantum field theory in curved space-time with matter quantization may explain the avoidance of big-bang singularity and emergent scenario as an alternative to it. Further, a complete cosmic evolution starting from inflationary era to the present accelerating can be described by suitable choice of continuous form of $f(R)$ function. The viability of $f(R)$-gravity theory as well as the form of $f(R)$ has been examined from the existing experimental results and observational data. Some constraints have been imposed on the form of $f(R)$. Finally, a thermodynamical analysis using the idea of unified first law of Hayward has been presented and it is possible to obtain a form of $f(R)$ from the validity of the Clausius relation. Therefore, in principle $f(R)$-gravity theory may be considered as an alternative to Einstein gravity for describing the whole cosmic evolution avoiding the big-bang singularity.   \\\\\\

 
 \section*{Acknowledgments} 
 Author DD thanks Department of Science and Technology
 (DST), Govt. of India for awarding Inspire research fellowship (File No: IF160067). SD acknowledge Science and Engineering Research Board (SERB),
 Govt. of India, for awarding National Post-Doctoral Fellowship
 (File No: PDF/2016/001435) and the Department
 of Mathematics, Jadavpur University where a part
 of the work was completed. SC thanks Science and Engineering Research Board (SERB) for awarding MATRICS Research Grant support (File No: MTR/2017/000407) and Inter University Center for Astronomy and Astrophysics (IUCAA), Pune, India for their
 warm hospitality as a part of the work was done during a visit.
 
\end{document}